# FUTURE TOP PHYSICS AT THE TEVATRON AND LHC

**A.P. Heinson**

Department of Physics, University of California,

Riverside, CA 92521–0413, USA

## Abstract

Top quarks are produced at the Tevatron proton-antiproton collider at Fermilab and at the Large Hadron (proton-proton) Collider at CERN in two ways: as quark-antiquark pairs, and singly. For each mode, the cross sections and future experimental yields are presented. I then discuss some precision measurements that can be made using the anticipated large data sets. These measurements include the top quark mass, width and branching fractions, the electroweak $Wtb$ coupling, and the CKM matrix element $V_{tb}$. Detailed studies of the top quark polarization and gluon radiation will also be possible, in addition to sensitive searches for $t\bar{t}$ resonances and rare decay modes.

# 1. Introduction

At hadron colliders, top quarks are produced via two independent mechanisms. The dominant mode is pair production of a top quark ($t$) and antiquark ($\bar{t}$) from $gg$ and $q\bar{q}$ fusion. This mode was used in the recent discovery of the top quark [1]. The second mode is single top production via the electroweak interaction: s-channel $q'\bar{q} \to t\bar{b}$ ($W^*$ single top); t-channel $q'b \to tq$ and $q'g \to tq\bar{b}$ (W-gluon fusion); and $bg \to tW$ ($tW$ production). This paper discusses some of the physics measurements we will be able to make using large numbers of top quark events at future hadron collider runs. The future run parameters for the Tevatron and LHC are given in Table 1. Shown for comparison is the already completed Tevatron "Run 1" where the two experiments DØ and CDF each collected a little over 100 pb$^{-1}$ of data. Both experiments are being extensively upgraded for Run 2 with the Main Injector in 1999, with 132 ns triggering capabilities. Tevatron Run 3, known as "TeV33" from the expected instantaneous luminosity of ~10$^{33}$ cm$^{-2}$s$^{-1}$, involves a modest proposed upgrade to the Tevatron, with a permanent magnet ring which will act as a recycler for the antiprotons. The TeV33 proposal has arisen from a year and a half long grass-roots study of possible high luminosity high $p_T$ physics opportunities that could be taken advantage of after Run 2. This study, the TeV-2000 project [2], contains many of the results presented in this paper.

Table 1  Future runs at the Fermilab Tevatron and CERN Large Hadron Collider (LHC).

| | Collider | | | |
|---|---|---|---|---|
| *Run Parameters* | Tevatron | | | LHC |
| Beams | $p\bar{p}$ | $p\bar{p}$ | $p\bar{p}$ | $pp$ |
| Run name | Run 1 | Run 2 (Main Injector) | Run 3 ("TeV33") | Run 1 |
| Energy $\sqrt{s}$ [TeV] | 1.8 | 2.0 | 2.0 | 14 |
| Run dates | 1992–1996 | 1999–2001 | 2003–2006 | 2005–2007 |
| Peak luminosity [cm$^{-2}$s$^{-1}$] | $2 \times 10^{31}$ | $2 \times 10^{32}$ | $10^{33}$ | $10^{32}$–$10^{33}$ |
| Integrated luminosity [fb$^{-1}$] | 0.1 | 2 | 30 | 10 |
| Beam crossing time [ns] | 3,500 | 396 → 132 | 132 | 25 |
| $<n>$ interactions / crossing | 1.9 | 6 → 2 | 9 | 0.2–1.9 |

The last column in Table 1 shows the expected running conditions at the start up of the Large Hadron Collider (LHC) at CERN. It is believed that the initial lower luminosity running during the first two years will be the period when most top physics will be done, since secondary vertex $b$-jet tagging will become very difficult at $\sqrt{s} = 14$ TeV in the full luminosity multiple interaction environment, with 19 interactions per bunch crossing at $10^{34}$ cm$^{-2}$s$^{-1}$ instantaneous luminosity.

# 2. Top Quark Production and Event Yields

The cross sections for top quark production at hadron colliders are shown in Table 2. These values are used to estimate the number of events produced in future runs, given in Table 3. At the Tevatron, there appears to be about 2.7 times as much $t\bar{t}$ as single top production, but recent next-to-leading order calculations of the single top cross sections [3] show that a $K$ factor of ~1.45 is needed, leading to a more realistic ratio of 2:1 for $t\bar{t}$:single top. At the LHC, the single top $K$ factor is ~1.32.

Table 2  Top quark cross sections at the Tevatron and LHC, showing the contributions to pair production rates from quarks and from gluons, and also the three separate single top production modes. The notation "+ C.C." means that the cross sections include the charge conjugate processes.

| $m_t$ = 170 GeV | Tevatron | LHC |
|---|---|---|
|  | $p\bar{p}$ | $pp$ |
| Energy $\sqrt{s}$ [TeV] | 2.0 TeV | 14 TeV |
| *Cross Sections* [pb] | | |
| top-antitop pairs[*] | 8.0 | 800 |
| $q\bar{q} \to t\bar{t}$ | 6.4 | 70 |
| $gg \to t\bar{t}$ | 1.6 | 730 |
| single top[#] | 3.0 | 400 |
| "W-gluon fusion" $q'b \to tq$, $q'g \to tq\bar{b}$; + C.C. | 1.8 | 253 |
| "$W^*$" $q'\bar{q} \to t\bar{b}$, $q'g \to t\bar{b}q$; + C.C. | 0.9 | 19 |
| $bg \to tW$, $q\bar{q} \to tW\bar{b}$, $gg \to tW\bar{b}$; + C.C. | 0.3 | 128 |

[*] The Tevatron value is a resummed next-to-leading order cross section from Laenen et al.[4], scaled from 1.8 TeV to 2.0 TeV [5]; and the LHC value is a leading order cross section at 14 TeV from Parke [6].
[#] Both the Tevatron and LHC values are leading order cross sections from calculations by Belyaev, Boos and Heinson [7].

Table 3  The numbers of top quark events produced per experiment at the Tevatron and LHC, and the predicted number of events reconstructed at the Tevatron. The top pair estimates are extrapolated from Run 1 experience. The single top numbers are from detailed Monte Carlo studies of signal and backgrounds. The notation "/b-tag" means at least one jet is b-tagged; "/b-b-tags" means two jets are b-tagged.

| $m_t$ = 170 GeV | Tevatron | | LHC |
|---|---|---|---|
|  | Run 2 | Run 3 | Run 1 |
| Integrated luminosity / experiment | 2 fb$^{-1}$ | 30 fb$^{-1}$ | 10 fb$^{-1}$ |
| *Produced* | | | |
| $t\bar{t}$ pairs | 16,000 | 240,000 | 8,000,000 |
| single top | 6,000 | 90,000 | 4,000,000 |
| *Reconstructed* | | | Tevatron S:B |
| $t\bar{t} \to ll + \geq 2\,\text{jets}$ | 193 | 2,895 | 5:1 |
| $t\bar{t} \to l + \geq 3\,\text{jets}/b$-tag | 1,374 | 20,610 | 3:1 |
| $t\bar{t} \to l + \geq 4\,\text{jets}/b$-$b$-tags | 607 | 9,105 | 12:1 |
| $q'b \to tq$, $\bar{t}q \to l + 2\,\text{jets}/b$-tag | 170 | 2,545 | 1:2.2 |
| $q'\bar{q} \to W^* \to t\bar{b}$, $\bar{t}b \to l + 2\,\text{jets}/b$-$b$-tags | 21 | 318 | 1:1.3 |

Table 3 shows the number of reconstructed top events from the two future Tevatron runs. One of the most important factors for high efficiency in finding top events is identifying, or "tagging", the jets originating from b-quarks. Tagging will be significantly improved in future runs by the addition to both DØ and CDF of large multilayer silicon microstrip detectors, with three-dimensional tracking and vertex reconstruction. These new detectors will cover the full interaction region, which will be reduced to $\sigma_z \approx 25\,\text{cm}$, and have geometrical acceptance of ~97% for b-jets from top decays. The b-tagging efficiency used here in the $t\bar{t}$ calculations assumes a 60% probability of tagging a fiducial b-jet. When this is combined with the efficiency to identify a nonisolated secondary lepton from the semileptonic decay of the *b*-hadrons (~15%), and the double tagging overlap fraction is subtracted, then the total tagging probability is 65% per *b*-jet from a 170 GeV top decay. The probability to tag at least one *b*-jet in a $t\bar{t}$ event is therefore 85%, and the probability to tag both *b*-jets is $\geq 42\%$. The double tag probability could be higher than 42% since the second tag criteria can be loosened, as backgrounds will be reduced after the first tag is made. For the single top calculations, a more conservative *b*-tag probability of 50% was assumed. Since the *W*+jets background is particularly severe for the lower jet multiplicity single top events, a nonzero mistag probability was included in the study. A value of 0.4% was chosen, based on experience with the CDF Run 1 silicon vertex detector, (conservative for future three-dimensional tagging), and this leads to an estimate that the background from *W*+*jj* events with a mistag (where *j* is a light quark or gluon jet) is as large as that from the three processes $W + b\bar{b}$, $W + c\bar{c}$ and $W + c$ with a real tag, combined.

## 3. Top Quark Mass

The top quark mass can be measured using several quasi-independent techniques, and the resulting measurements will then be combined for optimal precision. The methods include making kinematic fits to lepton+jets events (both single and double tagged), dilepton events, all-hadronic decays, and to single top events. Several techniques for kinematic fitting have been developed by DØ and CDF using Run 1 data, and the best of these methods has been used for this study. In addition, the top quark mass can be measured by fitting the mean *b*-hadron lifetime. The systematic error on this measurement will be almost independent of that from kinematic fitting, as it will not depend on the jet energy scale.

For the TeV-2000 report, detailed studies of the systematic errors on the top mass measurement were made, to establish which sources scale with statistics, and which improve less rapidly. These studies serve as a benchmark for the precision obtainable across the top physics program. Results for the most sensitive channel are shown in Table 4. For comparison, the current preliminary top mass measurements from Run 1 use ~30 lepton+jets events each, and get a statistical error of 6–15 GeV and a systematic error of 7–10 GeV, giving a total error of 5–10%.

Table 4    The estimated error on the top quark mass from Tevatron Runs 2 and 3, using the most sensitive *l*+jets channel with at least one tagged *b*-jet. The values come from detailed studies of how the Run 1 mass measurement systematic errors scale with increased statistics, and include new methods of energy scale calibration.

| $m_t = 170$ GeV | Tevatron | |
|---|---|---|
| | Run 2 | Run 3 |
| Integrated luminosity / experiment | 2 fb$^{-1}$ | 30 fb$^{-1}$ |
| $t\bar{t} \rightarrow l+ \geq 4\,\text{jets}/b\text{-tag}$    # of events | 1,216 | 18,240 |
| statistical error    [MeV] | 1,000 | 300 |
| systematic error    [MeV] | 2,600 | 700 |
|     jet energy scale | 2,300 | 600 |
|     *b*-tagging bias | 500 | 100 |
|     background shape | 1,100 | 300 |
| Total error on top quark mass | 2.8 GeV | 800 MeV |
| | 1.6 % | 0.5 % |

Techniques to constrain the dominant jet energy scale error include: reconstructing $Z \to e^+e^- + 1$ jet events; using the $W \to$ jet-jet decays within the top decays, which are quark jets uncontaminated by $b$-jets; and using $Z \to b\bar{b}$, or $WZ$ with $Z \to b\bar{b}$, to calibrate the calorimeter response to $b$-jets.

The top mass measurement may be combined with the $W$ mass measurement from the same experiment to obtain limits on the Higgs boson mass. From Run 3 with 30 fb$^{-1}$ per experiment, the $W$ mass will be measured to 15 MeV [1] (0.02%) at the Tevatron, and when combined with the $< 1$ GeV (0.5%) precision on the top mass, the Higgs mass will be constrained to within 30% of itself. This level of precision will be extremely useful for the LHC experiments ATLAS and CMS when they start to look for standard model and supersymmetric Higgs bosons, particularly if the Higgs is light. The LHC experiments should obtain a similar precision on their top mass measurements as that estimated for the Tevatron.

## 4. Top Quark Cross Section and Resonance Searches

Measurements of the top quark cross sections at the Tevatron and LHC are complementary, since production at the Tevatron is dominated by valence quark annihilation, whereas at the LHC it is predominantly via gluon fusion. Table 5 shows the precisions obtainable on $t\bar{t}$ pair and single top cross sections at the Tevatron. The sensitivity is limited either by low statistics, for the Run 2 single top measurements, or by the 5% error on the luminosity for all other results. For comparison, the current experimental error on the $t\bar{t}$ cross section at the Tevatron is 25-35%. The experimental values can be compared with higher order theoretical calculations, and any excess or deficit seen could be generated by new physics. For instance, a heavy resonance decaying into $t\bar{t}$ pairs will boost the cross section, and produce a peak in the $t\bar{t}$ invariant mass distribution. Table 5 shows the sensitivity to find such a peak at the $5\sigma$ sensitivity level for a topcolor $Z'$ [8]. Various models for dynamical electroweak symmetry breaking predict many heavy color-singlet and color-octet states. The Tevatron should be sensitive to either type, whereas the gluon fusion dominated LHC will be insensitive to spin one color singlets such as the $Z'$.

Table 5  The top quark cross section precision from Tevatron Runs 2 and 3. The dominant contribution to the error in Run 3 is the 5% error on the integrated luminosity. Run 2 single top studies are statistics limited. The estimated discovery reaches for heavy particles decaying into $t\bar{t}$ pairs are given for a possible $Z'$ from topcolor-assisted technicolor, with either a narrow or wide resonance width.

| $m_t = 170$ GeV | | Tevatron | |
|---|---|---|---|
| | | Run 2 | Run 3 |
| Integrated luminosity / experiment | | 2 fb$^{-1}$ | 30 fb$^{-1}$ |
| *Cross Section Precision* | | | |
| top-antitop pairs | | 8 % | 6 % |
| all single top | | 25 % | 7 % |
| just $q'\bar{q} \to W^* \to t\bar{b}, \bar{t}b$ | | 27 % | 8 % |
| *Resonance Search Sensitivity** | | | |
| $\sigma \times B(X \to t\bar{t})$ | $\Gamma_X = 1.2\% \, m_X$ | < 80 fb | < 5 fb |
| $m_X$ | $\Gamma_X = 1.2\% \, m_X$ | > 920 GeV | > 1150 GeV |
| $\sigma \times B(X \to t\bar{t})$ | $\Gamma_X = 10\% \, m_X$ | < 600 fb | < 34 fb |
| $m_X$ | $\Gamma_X = 10\% \, m_X$ | > 560 GeV | > 820 GeV |

\* The limits shown are at the $5\sigma$ sensitivity level (99.999943% CL) appropriate for discovery.

## 5. Top Quark Width

The width of the top quark is a fundamental parameter of the standard model, so far not measured. Theoretical calculations give a value of 1.4 GeV for a top mass of 170 GeV, which corresponds to a lifetime of $0.47 \times 10^{-24}$ s.

The Tevatron will provide a uniquely powerful facility for measuring the top quark width, for the following reason. The branching ratios of top decaying into *Wb/Wq* and *Wb/Xb* can be measured at any facility that produces a large number of $t\bar{t}$ events, but measuring the partial width $\Gamma(t \to Wb)$ can best be done using the s-channel $W^*$ single top production rate [9], and this process is only accessible at a $p\bar{p}$ collider such as the Tevatron. The reason this process is so powerful is that all sources of systematic error in the theoretical calculation of the cross section are understood to be small and can be accounted for, or measured, using for instance the similar Drell-Yan process. This is not the case if *W*-gluon fusion were to be used, when the lack of precise knowledge of the gluon distribution within the proton has a detrimental effect on the final result. Table 6 shows the various measurements that need to be made in order to calculate the total top quark width.

Table 6  Top quark branching ratios and the partial and total width measurements from Tevatron Runs 2 and 3. These results include estimates of statistical and systematic errors.

| $m_t = 170$ GeV | Tevatron | |
|---|---|---|
| | Run 2 | Run 3 |
| Integrated luminosity / experiment | 2 fb$^{-1}$ | 30 fb$^{-1}$ |
| *Branching Ratios Precision* | | |
| $B(t \to Wb)/B(t \to Wq)$ * | 2 % | 0.5 % |
| $B(t \to Wb)/B(t \to Xb)$ # | 7 % | 2 % |
| *Top Width Precisions* | | |
| $\Gamma(t \to Wb)$ † | 27 % | 8 % |
| $\Gamma(t \to X)$ ‡ | 28 % | 9 % |

\* This measurement is made using three methods: (a) the ratio of single *b*-tags to double *b*-tags in $t\bar{t} \to l +$ jets events; (b) the number of *b*-tags in $t\bar{t}$ dilepton events; and (c) the ratio of double *b*-tags in one jet of $t\bar{t}$ events to the number of $t\bar{t}$ events with both jets tagged.
\# This comes from the ratio of the number of $t\bar{t}$ dilepton events to the number of $t\bar{t}$ lepton+jets events.
† The partial width of the top quark to decay to *Wb* is directly proportional to the cross section for single top production. The error is the sum in quadrature of the measured cross section error and the theory cross section error (the measurement error dominates, mostly from the 5% error on the luminosity).
‡ The total top quark width is obtained by combining the previous three measurements.

The LHC experiments will probably not be able to make a better measurement of the total top quark width, despite the very large statistics available, as it is unlikely that the total error on the theoretical cross section from the gluon distribution function will be known to better than ~10% for the *W*-gluon fusion process. In addition at the LHC, it has been shown that it will not be possible to separate out the $W^*$ single top process, due to its low cross section and the high backgrounds [9]. It should also be noted that at a future linear $e^+e^-$ collider, the cross section for single top production is minute [10]. Discussion of a $t\bar{t}$ threshold scan in the papers mentioned in Ref. 11 suggests that measuring the top quark width this way will be a very difficult measurement, requiring mature understanding of the linear collider performance.

## 6. Wtb Coupling and $V_{tb}$

The *Wtb* coupling can only be probed directly using top quark events. It is a *V-A* coupling, proportional to the CKM matrix element $V_{tb}$, and is given by (for example, in the unitarity gauge):

$$\Gamma = \frac{eV_{tb}}{2\sqrt{2}\sin\vartheta_W}\left[\gamma_\mu(1-\gamma_5)\right] \qquad (1)$$

In the standard model, top quarks decay only to left-handed or longitudinally polarized *W* bosons, with the ratio given by:

$$\frac{W_{\text{long}}}{W_{\text{left}}} = \frac{m_t^2}{2m_W^2} = 2.24 \qquad (2)$$

for $m_t = 170$ GeV. Nonuniversal top couplings could appear as a departure from the branching fraction of top to $bW_{\text{long}}$ (69%). The *W* polarization is accessible experimentally by studying the neutrino decay direction from a leptonic decay [12], or via the charged lepton angular distribution [13], presented here. In the laboratory frame, the cosine of the polar angle of the lepton is given by:

$$\cos\vartheta_l^* \approx \frac{2m_{lb}^2}{m_{lb\nu}^2 - m_W^2} - 1 \qquad (3)$$

where $\vartheta_l^*$ is the angle between the lepton and the *+z* axis in the rest frame of the *W* boson, and the *+z* axis is chosen to be opposite to the *b*-quark direction in the rest frame of the top quark.

Using this relation, and fitting to the fraction of longitudinal and left-handedly polarized *W* bosons from top decay, one can measure the branching fractions to the precision shown in Table 7. There should be no right-handed *W* component and therefore the sensitivity to such a decay is rather good. A sample of 10,000 top events could be used to detect a 5% fraction of $W_{\text{right}}$ decays with $5\sigma$ sensitivity.

The CKM matrix element $V_{tb}$ can be probed using both top pairs and single top events. When a *t*-quark and a $\bar{t}$-antiquark are produced together, they have almost no net transverse polarization, but their spins are correlated [14]. At the Tevatron, ~70% of the $t\bar{t}$ pairs have opposite helicity, whereas at the LHC, ~65% of the $t\bar{t}$ pairs have the same helicity. Studying the angular distributions of the charged lepton and the lightest quark leads to a lower limit on $|V_{tb}|$. The single top production cross section is directly proportional to $|V_{tb}|^2$, so a measurement of the single top rates, (in particular $W^*$ production where the theory error is small and well-known), leads to high precision on $|V_{tb}|$, as shown in Table 7.

Table 7  Top quark branching fractions to left-handed and longitudinally polarized *W* bosons, and the precision obtainable on $|V_{tb}|$ from single top, in Tevatron Runs 2 and 3. These results include statistical and systematic errors.

| $m_t = 170$ GeV | Tevatron | |
| --- | --- | --- |
|  | Run 2 | Run 3 |
| Integrated luminosity / experiment | 2 fb$^{-1}$ | 30 fb$^{-1}$ |
| *Branching Fractions Precision* | | |
| $B(t \to bW_{\text{long}})$ | 4.6 % | 1.2 % |
| $B(t \to bW_{\text{right}})$ | 1.8 % | 0.5 % |
| $|V_{tb}|$ *Precision* | | |
| From $W^*$ single top rate | 14 % | 4 % |

## 7. Top Quark Radiation and Rare Decays

Within the standard model, the top quark can radiate gluons and/or photons. Many theoretical studies of the angular distributions, interferences, energy spectra and so on of such radiation exist [15] and precision measurements of such final states will be possible with the large data sets from future runs at the Tevatron and LHC. These studies will facilitate precision tests of QCD.

If the top quark mass is > ~176 GeV, then it will lie above the threshold for decaying into $WZb$ [16]. This standard model decay can be found by looking for extra lepton pairs, or $b$-tagged jets which reconstruct to produce the $Z$ boson invariant mass. Such events, although rare, would be useful for calibrating the top quark mass measurement. An enhanced rate would indicate new physics.

If the neutral standard model Higgs boson is light enough, then it too could be radiated during the decay of the heavy top quark, giving $t \to WHb$ [16]. Decays with extra $b$-tagged jets which reconstruct to give a peak in an invariant mass spectrum would allow study of the Higgs boson even without sufficient beam energy for pair production.

Beyond the standard model, there are many predicted decay channels for the top quark. Sensitivity levels for searches at the Tevatron for three possible decays are shown in Table 8. Decays into a charm or up quark plus a neutral particle such as a photon, gluon, $Z$ or Higgs boson are predicted within the standard model, but at vanishingly low rates. Some extensions to this model boost production [17].

Table 8    Sensitivities for searches at the upgraded Tevatron for rare decays of top quarks (flavor changing neutral currents, FCNC, and supersymmetry, SUSY). Limits are given at the 95% confidence level, and include both statistical and systematic errors.

| $m_t = 170$ GeV | | Tevatron | |
|---|---|---|---|
| | | Run 2 | Run 3 |
| Integrated luminosity / experiment | | 2 fb$^{-1}$ | 30 fb$^{-1}$ |
| *Rare Decays Search Limits* | | | |
| $t \to c/u + \gamma$ | FCNC | < 0.3 % | < 0.02 % |
| $t \to c/u + Z$ | FCNC | < 2 % | < 0.1 % |
| $t \to H^+ b$ | SUSY | < 11 % | < 3 % |

Supersymmetry predicts many new decays of the top quark [18], in particular to a charged Higgs boson and a $b$-quark [19]. The indirect limit on $t \to$ non-$W$ states in Run 3 of the Tevatron, (from the ratio of the $t\bar{t}$ dilepton to lepton+jets cross sections), coupled with the $b \to s\gamma$ measurement by CLEO [20], will lead to the discovery of the charged Higgs boson if it is lighter than the top quark, for any value of $\tan\beta$. Other supersymmetric decays of the top quark include $t \to \tilde{t}\tilde{Z}$ [21]. With large event samples of $t\bar{t}$ pairs, searches for a wide variety of rare and nonstandard decays will be possible. One of the two tops can be unambiguously identified using the leptonic decay of the $W$ boson and a tagged $b$-jet, and then the other top decay can be carefully examined to search for any of the many exotic predicted final states.

## 8. Summary

The TeV-2000 project has studied a wide range of top physics that can be done at future Tevatron runs. The two experiments DØ and CDF are being upgraded for the next collider run, and it is now known that there is a significant broad-based program of top physics for the Tevatron into the next millenium. Some of the measurements possible at the Tevatron, such as the top quark width, $|V_{tb}|$, and searches for certain high mass resonances can be done with greater precision than at any other collider, since they take advantage of processes generated from the valence quarks and antiquarks that dominate at a proton-antiproton machine.

The LHC will produce an extremely large number of top quarks, both in pairs, and via single top *W*-gluon fusion and *tW* production. Measurement of the cross sections at the LHC is complementary to measurements made at the Tevatron, because of the different collision energy and production mechanisms. The precision obtainable on the top quark mass should be similar to that achievable at Fermilab. The very high statistics at the LHC should enable first observation of some of the rare decay modes of the top quark.

**Acknowledgments**

I would like to thank the organizers of the XXXIst Moriond QCD Meeting, especially J. Tran Thanh Van, for one of the best conferences I have attended.

The top physics studies presented in this paper are the work of many people, mainly for the TeV-2000 project. I would like to thank them for a very enjoyable collaboration. The authors of the top physics chapter of the TeV-2000 report are: Dan Amidei, Phil Baringer, Nathan Eddy, Ulrich Heintz, Soo Bong Kim, Tom LeCompte, Tony Liss, Meenakshi Narain, Kirsten Tollefson, Steve Vejcik, Brian Winer and Dave Winn, from the DØ and CDF Collaborations. The instigators of TeV-2000 and editors of the grand report are Dan Amidei and Chip Brock, and their foresight and vision have made this possible.